\documentclass[twoside]{article}
\usepackage{qic}
\usepackage{subcaption}
\usepackage{amsfonts}
\usepackage{amsmath}
\usepackage[linesnumbered, ruled,vlined]{algorithm2e}
\usepackage{graphicx,booktabs,array}
\usepackage[braket, qm]{qcircuit}
\usepackage{xcolor} 
\usepackage{tikz}
\usepackage{diagbox}
\usetikzlibrary{fit}
\usepackage{hyperref}
\hypersetup{urlcolor=violet, citecolor=violet, linkcolor=violet, colorlinks=true}

\SetKwFunction{RowCol}{RowCol}
\SetKwFunction{SteinerGauss}{SteinerGauss}

\newcommand{\eref}[1]{(\ref{#1})}
\newcommand{\sref}[1]{section~\ref{#1}}
\newcommand{\fref}[1]{figure~\ref{#1}}
\newcommand{\tref}[1]{table~\ref{#1}}

\newcommand{\Sref}[1]{Section~\ref{#1}}

\newcommand{\mr}{\ms \hline \ms}
\newcommand{\ms}{\noalign{\vspace{1pt}}}

\colorlet{gray}{gray!50}

\begin{document}
\setlength{\textheight}{8.0truein}    %FOR 2ND PAGE ONWARDS

\runninghead{Linear circuit synthesis using weighted Steiner trees}
            {Nir Gavrielov, Alexander Ivrii, and Shelly Garion }

\normalsize\textlineskip
\thispagestyle{empty}
\setcounter{page}{1}

%\copyrightheading{Vol.}{No.}{Year}{Page Nos.}
% \copyrightheading{0}{0}{2003}{000--000}

\vspace*{0.88truein}

\alphfootnote

\fpage{1}

\centerline{\bf
%%%%%%%%%%%%%%%%%%%%%
%Put in titiles here
%%%%%%%%%%%%%%%%%%%%%
LINEAR CIRCUIT SYNTHESIS USING WEIGHTED STEINER TREES}
\vspace*{0.035truein}
\vspace*{0.37truein}
\centerline{\footnotesize
%%%%%%%%%%%%%%%%%%%%%%%%%%%%%%%%%%%%
%put authors' name and address here
%%%%%%%%%%%%%%%%%%%%%%%%%%%%%%%%%%%%
NIR GAVRIELOV }
\vspace*{0.015truein}
\centerline{\footnotesize\it Racah Institute of Physics, The Hebrew University of Jerusalem, Givat Ram}
\baselineskip=10pt
\centerline{\footnotesize\it Jerusalem 91904, Israel}
\vspace*{10pt}
\centerline{\footnotesize 
ALEXANDER IVRII, SHELLY GARION}
\vspace*{0.015truein}
\centerline{\footnotesize\it IBM Quantum, IBM Research Israel, Haifa University Campus, Mount Carmel}
\baselineskip=10pt
\centerline{\footnotesize\it Haifa 3498825, Israel}
\vspace*{0.225truein}

\vspace*{0.21truein}

%% \abstracts{first paragraph}{second paragraph}{third paragraph}
%% If there is only one paragraph, just keep the second and third empty 
%% like the following one 
\abstracts{
%%%%%%%%%%%%%%%%%%%%
% put abstract here
%%%%%%%%%%%%%%%%%%%%
CNOT circuits are a common building block of general quantum circuits. The problem of synthesizing and optimizing such circuits has received a lot of attention in the quantum computing literature. This problem is especially challenging for quantum devices with restricted connectivity, where two-qubit gates can only be placed between adjacent qubits. The state-of-the-art algorithms for optimizing the number of CNOT gates are heuristic algorithms that are based on Gaussian elimination and that use Steiner trees to connect between different subsets of qubits. In this article, we suggest considering \emph{weighted} Steiner trees, and we present a simple low-cost heuristic to compute weights. The simulated evaluation shows that the suggested heuristic is almost always beneficial and reduces the number of CNOT gates by up to 10\%.}{}{}

\vspace*{10pt}

\vspace*{1pt}\textlineskip    %) USE THIS MEASUREMENT WHEN THERE IS
   %) A SECTION HEADING
%\vspace*{-0.5pt}
%\noindent
%%%%%%%%%%%%%%%%%%%%%%%%%%%%%%%%
%put the text of the paper here
%%%%%%%%%%%%%%%%%%%%%%%%%%%%%%%%
\section{Introduction}
\label{sec:introduction}
Quantum circuit synthesis is a process of constructing a quantum circuit that implements a given unitary operator and can be executed on a given quantum device while minimizing the number of gates or the depth of the circuit. Quantum devices differ in terms of \emph{supported gate set} and \emph{connectivity}. For instance, superconducting quantum devices support single-qubit rotations and the two-qubit CNOT gates; moreover, the CNOT gates can only be placed between the ``connected" qubits. 
Hence, placing an entangling two-qubit gate on non-connected qubits requires finding an optimal route between these two qubits and placing SWAP gates between all pairs of adjunct qubits in this route, where each SWAP gate can be implemented using three CNOT gates \cite{routing_swaps}. 
As the two-qubit gates are significantly noisier than single-qubit gates, the goal of quantum circuit synthesis is usually to optimize the number of CNOT gates or the CNOT depth of the constructed circuit.

CNOT circuits (that is, circuits that only consist of CNOT gates) appear as common building blocks of general quantum circuits. For instance, they are common subcircuits produced by variational-quantum-eigensolver (VQE) algorithms used for quantum chemistry, quantum simulation and quantum optimization (see, e.g.~\cite{VQE_ansatz}). CNOT circuits also appear when studying Clifford operators, \cite{DBLP:journals/corr/MaslovR17} shows that any Clifford operator can be implemented in the form $-CX-CZ-P-H-P-CZ-CX-$, which includes two $CX$-layers (that is, two CNOT subcircuits). Thus, it should not be surprising that the problem of synthesizing and optimizing CNOT circuits has received a lot of attention in the quantum computing literature. 

The problem of synthesizing a CNOT-circuit of width n can be viewed as the problem of reducing a matrix in $GL_2 (n)$ to the identity matrix using Gaussian elimination (see \sref{sec:preliminaries} for details), with the additional restriction that row operations can only be performed between rows corresponding to connected qubits. Two algorithms for this task are described in~\cite{Nash_2020, SteinerGauss, RowCol}, which are both heuristic algorithms that aim to minimize the number of CNOT gates. The
\SteinerGauss algorithm~\cite{Nash_2020, SteinerGauss} is based on reducing first a matrix in $GL_2(n)$ to a lower-triangular matrix and then to the identity matrix. The \RowCol algorithm~\cite{RowCol} is based on eliminating one qubit at a time. Crucially, both works rely on the computation of Steiner trees between various sets of qubits to optimize the number of row operations for each basic step of the algorithm (e.g., eliminating all non-diagonal $1$s in the given column of a matrix). 
In all of these works, Steiner tree optimization aims to minimize the number of edges needed to connect a given subset of nodes. As the problem is NP-hard, in practice approximate Steiner tree computation algorithms are used.

In general, at each step of the algorithm, multiple different optimal Steiner trees can be found, with each tree leading to a different sequence of row operations and hence to a different matrix obtained by executing these row operations. This work aims to improve the above algorithms based on the intuition that it is beneficial to choose the tree that brings the matrix ``closer'' to the target, the identity matrix. To this extent, we suggest assigning weights to the edges of the connectivity graph and present a heuristic for doing so. This heuristic has a low computational cost and can be easily used with any standard approximate Steiner tree computation tool. Similar intuition and a related heuristic were described in~\cite{Vandaele_2022} in the context of phase polynomial synthesis.
The experimental evaluation shows that the suggested heuristic is beneficial and reduces the number of CNOT gates.

In addition, we improve the CNOT circuit depth compared to the \RowCol and \SteinerGauss algorithms mentioned above. We note that for certain connectivity maps between the device qubits, there are better known estimations. For all-to-all connectivity of the $n$ qubits, where every pair of qubits is connected, the depth of the CNOT circuit is bounded by $n + O(\log ^ 2 (n))$
\cite{MZdepth, Brugdepth}. For linear nearest neighbor connectivity of the $n$ qubits the depth is bounded by $5n$ \cite{KMS}. 
In \cite{RowCol} it was shown that for a two-dimensional grid connectivity of the $n$ qubits, the depth is $O(n)$ with $n^2$ ancillas. Recently,~\cite{shallow} extended the approach of~\cite{KMS} to handle blocks of qubits arranged in a line, and in particular proved that the depth of the CNOT circuit on the $n$-qubit grid is bounded by $4n$ (without using ancilla qubits).
In some of the current quantum devices, the connectivity between the qubits
is a two-dimensional heavy hexagon, into which one can embed a line or a sequence of blocks arranged in a line that includes all or almost all of the qubits, obtaining an effective bound on the CNOT circuit depth.
However, such a bound is not known for all possible quantum device connectivity maps, and our results provide a heuristic algorithm that improves the depth in general.

This paper is organized as follows. In \Sref{sec:preliminaries}, we describe the CNOT circuit synthesis problem and the \SteinerGauss and \RowCol algorithms. In \Sref {sec:heuristic}, we describe the weight assignment heuristic and illustrate it with an example. \Sref{sec:experiments} contains the experimental evaluation. \Sref{sec:conclusion} concludes the paper.

\section{Preliminaries}
\label{sec:preliminaries}
\paragraph{Linear circuits} 

Quantum circuits consisting only of CNOT gates are known as \emph{linear circuits} or \emph{linear functions}~\cite{PMH}. The CNOT gate performs a controlled-not operation on two qubits:
$CNOT(c,t) = (c,t\oplus c)$, with $c$ called the
\emph{control} qubit and $t$ called the \emph{target} qubit. 
Hence, a CNOT gate between two qubits corresponds to a linear reversible function $f: \mathbb{F}_2^2 \rightarrow \mathbb{F}_2^2$, which is a $2 \times 2$ binary invertible matrix. 
The CNOT gate is also universal for linear reversible circuits, therefore any linear reversible function, i.e., any $n\times n$ matrix in the $GL_2 \left(n\right)$ group, can be implemented using only CNOT gates. A single CNOT circuit gate, controlled by wire $i$ and acting on wire $j\neq i$, can be represented by the elementary matrix $E_{ij}$ (equal to the identity matrix with component $\left(i,j\right)$ flipped to 1). Composing a succession of these operations one obtains a $n \times n$ \textit{parity matrix}, representing a n-qubit linear circuit. Row~$i$ of the matrix contains the parity output of qubit~$i$, sum of those qubits with index~$j$ such that the entry $\left(i,j\right)$ of the matrix equals~$1$. Thus, given a random reversible operator $M$ we look for a sequence of $m$ elementary matrices that satisfy 
\begin{equation}
\left(\prod_{k=1}^{m}E_{i_{k},j_{k}}\right)M=I_{n}\Leftrightarrow\prod_{k=m}^{1}E_{i_{k},j_{k}}=M
\end{equation} using $E_{ij}^{-1}=E_{ij}$. A circuit containing CNOT operations corresponding to the sequence of elementary matrices $\left(E_{i_{k},j_{k}}\right)_{k=m}^{1}$ implements the function $M$. For example, the linear circuit in \fref{fig:synth_example} can be represented by
\begin{equation} 
\label{eq:synth_example} 
\begin{pmatrix} 
1 & 1 & 0 & 0\\
0 & 1 & 0 & 0\\
0 & 0 & 1 & 0\\
0 & 0 & 0 & 1 
\end{pmatrix}
\begin{pmatrix} 
1 & 0 & 0 & 0\\
0 & 1 & 1 & 0\\
0 & 0 & 1 & 0\\
0 & 0 & 0 & 1 
\end{pmatrix}
\begin{pmatrix} 
1 & 0 & 0 & 0\\
0 & 1 & 0 & 0\\
0 & 0 & 1 & 1\\
0 & 0 & 0 & 1 
\end{pmatrix}
\begin{pmatrix} 
1 & 0 & 0 & 0\\
1 & 1 & 0 & 0\\
0 & 0 & 1 & 0\\
0 & 0 & 0 & 1 
\end{pmatrix}
\begin{pmatrix} 
1 & 0 & 0 & 0\\
0 & 1 & 0 & 0\\
0 & 0 & 1 & 0\\
0 & 1 & 0 & 1 
\end{pmatrix}
= 
\begin{pmatrix} 
0 & 0 & 1 & 1\\
1 & 0 & 1 & 1\\
0 & 1 & 1 & 1\\
0 & 1 & 0 & 1 
\end{pmatrix} 
\end{equation}

Since two-qubit gates are prone to hardware errors, a synthesis algorithm will aim to reduce the total amount of operations, known as the circuit size. A second key metric is the resulting circuit depth, corresponding to the number of timestamps required to execute the circuit, assuming that independent gates are performed simultaneously. 
\begin{figure}[h]
\[
\begin{array}{c}
     
\Qcircuit @C=1.0em @R=0.8em @!R { \\
\nghost{{q}_{0} :  } 
& \lstick{{q}_{0} :  } &\qw &\qw &\qw &\ctrl{1} &\qw &  \qw &\qw & \targ &\qw \\ 
\nghost{{q}_{1} :  } 
& \lstick{{q}_{1} :  } &\qw &  \ctrl{2} & \ustick{G_2} \qw & \targ  &\qw &  \targ &\ustick{G_5} \qw & \ctrl{-1} & \qw\\
\nghost{{q}_{2} :  } 
& \lstick{{q}_{2} :  } &\ustick{G_1} \qw & \qw &\qw &  \targ & \ustick{G_4} \qw &  \ctrl{-1} & \qw & \qw & \qw \\
\nghost{{q}_{3} :  } 
& \lstick{{q}_{3} :  } &\qw& \targ & \ustick{G_3} \qw &  \ctrl{-1} &\qw &\qw & \qw & \qw & \qw \\
}
\end{array}
\]
\fcaption{\label{fig:synth_example} A linear circuit with depth $4$ and $5$ CNOT gates, represented by the 5 elementary operations in Eq.~(\ref{eq:synth_example}), in which and in left-to-right order $G_5$ corresponds to the first matrix and $G_1$ to the last. }
\end{figure}

\paragraph{Steiner Trees} A coupling map of a quantum processor is usually represented using an undirected graph $G=(V,E)$ where $V$ is a set of vertices (qubits) and the edges E denote the coupling between adjacent qubits. Such graph has at most one edge connecting each pair of nodes (not a ``multi-graph'') and no self-loops, i.e. edges with the same node on both ends. The edges of a graph can be given a numerical weight by some rule, making it a weighted graph. A tree is an undirected graph in which any two vertices are connected by a unique path - a connected graph with no loops. A spanning tree of a connected graph is a tree subgraph that includes all the vertices of~$G$. Each graph may contain many spanning trees and those with minimal overall edge weight are called minimal spanning trees~(MST). Given a graph $G=(V,E)$ and a set of vertices $S$, a \textit{Steiner Tree} $T=(V_T,E_T)$ is a minimal weight tree subgraph that contains all the vertices in~$S$. The nodes in $S$ are often called \textit{terminals} and those in $V_T\setminus S$ are known as \textit{Steiner nodes}. The problem of finding an optimal Steiner tree in an arbitrary graph is known to be NP-hard \cite{Karp1972}. Our work uses the rustworkx graph package~\cite{Treinish_2022}, therefore to find a Steiner tree we used their method based on the $(2-2/|S|)$ approximate algorithm from \cite{MEHLHORN1988125}.
A special kind of Steiner tree used is a \textit{decreasing Steiner tree}, in which each node is larger than its children according to a certain input ordering of the vertices.

\paragraph{Steiner tree based synthesis of linear circuits} 
We briefly describe the state-of-the-art algorithms \SteinerGauss and \RowCol for the synthesis of linear circuits in quantum devices with restricted connectivity. The algorithms start with a linear invertible binary matrix and aim to optimize the number of row operations (adhering to the connectivity of the device) that reduce the matrix to the identity or to a permutation matrix.

The \SteinerGauss~\cite{Nash_2020, SteinerGauss} algorithm first reduces the original matrix to an upper triangular matrix. The matrix $M$ is brought into the upper triangular form column after column. When processing a column $i$, one considers the set $S = \{i\} \cup \{k  |\ k>i \ and\ M_{k, i}=1\}$ corresponding to the diagonal entry of $M_{i, i}$ together with the column's $1$-entries below the diagonal. These entries are then connected using an (approximated) Steiner tree, which is then used to perform a sequence of row operations that result in $M_{i, i}=1$ and $M_{k, i} = 0$ for all $k>i$. In the second step, the matrix $M$ is reduced to the identity matrix. This is achieved by transposing the matrix (to become lower-triangular) and applying the algorithm from the previous step with one important change required to preserve the lower-triangular form: only \emph{decreasing} Steiner trees can be used. 

The \RowCol algorithm~\cite{RowCol} follows a somewhat different matrix simplification strategy. This algorithm is presented as Algorithm~\ref{alg:wRC} (without the gray highlighted text).
%We will later use this algorithm to explain our method. 
\RowCol processes one node of $V$ after another, each time fully simplifying both the relevant column and the relevant row of $M$. Given a node $i\in V$, in lines $3$--$6$ of the algorithm the diagonal element $M_{i, i}$ is turned to $1$ and all non-diagonal elements in column $i$ are turned to $0$. This is similar to the first step of \SteinerGauss, except that all the $1$ entries in column $i$ are considered. In lines $8$--$11$ all the non-diagonal elements in row $i$ are turned to $0$. 
To do so, the algorithm finds a linear combination of rows that is equal to the target row plus the relevant unit vector. 
These vertices are added to the target row, using Steiner trees to guide how the rows are added. At this point
$M_{i, i} = 1$ and $M_{i, k} = M_{k, i} = 0$ for all $k\ne i$. On line $12$ the vertex $i$ is removed from the graph, and the process continues until the graph becomes empty.

\SetKwComment{Comment}{/* }{ */}

\begin{algorithm}[ht]
\DontPrintSemicolon\SetAlgoLined
\SetKwInOut{Input}{Input}
\SetKwInOut{Output}{Output}
\SetKwFunction{UpdateEdgeWeight}{UpdateEdgeWeight}
\Input{Integer n, matrix $\textbf{M}\in\mathbb{F}_2^{n\times n}$, graph $\left(V,E\right)$ where $|V| = n$, \\
\colorbox{gray!40}{weight function $w: \mathbb{F}_2^{n\times n} \times E\rightarrow\mathbb{R}^{\geq0}$}}

\Output{Row additions to transform \textbf{M} into \textbf{I}}
%\SetKwProg{myalg}{RowCol}{}{}
\def\HiLi{\leavevmode\rlap{\hbox to \hsize{\color{gray!40}\leaders\hrule height .8\baselineskip depth .5ex\hfill}}}
%\myalg{}{
\For {$i\in V$ which is not a cut vertex}{
  \HiLi {$\UpdateEdgeWeight\left(\textbf{M}, G, w\right)$}\;
  $S = \left\{j| \textbf{M}_{ji}\neq 0\right\} \cup \left\{i\right\}$\;
  Find a tree T containing $S\subseteq V$ in G\;
  Postorder traverse T from i. When reaching j with parent k, add row j to row k if $\textbf{M}_{ji}=1$ and $\textbf{M}_{ki}=0$\;
  Postorder traverse T from i, add every row to its children when reached\;
  \HiLi {\UpdateEdgeWeight $\left(\textbf{M}, G, w\right)$}\;
  Let $S^{\prime}\subseteq V$ that $\sum_{j\in S^{\prime}}\textbf{M}_j = \textbf{M}_i + e_i$\;
  Find a tree $T^{\prime}$ containing $S^{\prime}\cup \{i\}$\;
  Preorder traverse $T^{\prime}$ from i. When reaching $j\notin S^{\prime}$, add the j-th row to its parent\;
  Postorder traverse $T^{\prime}$ from i, add every row to its parent when reached\;
  Delete i from graph G\;
  }
\caption{Weighted RowCol, weighted edges optimization extension of the RowCol algorithm. Similar adjustment can be used in SteinerGauss or any other Steiner tree based algorithm.}
\label{alg:wRC}
\end{algorithm}

\section{Method}
\label{sec:heuristic}
 We propose a heuristic method for weighting the edges of the coupling graph during a synthesis algorithm, which leads to a reduced total number of CNOT gates in the transpiled circuit. In the linear function synthesis task, one aims to transform a general binary matrix to the identity using row operations. Since the atomic operator in the process is the addition of two binary rows of the matrix, we look for a function that given two such vectors assigns a scalar weight. A successful choice of a function of this kind will result in edges representing the number of CNOT gates added upon their choice, and the optimal Steiner tree will include edges that minimize the total number of gates added in the current step of the algorithm. However, such a function is hard to find, as there is no exact quantitative measure connecting the linear function matrix to the number of gates in the final circuit. Furthermore, after obtaining an optimal tree, the mentioned algorithms apply many additions in both directions, that is, parent-to-child and vice versa, actions that have to be taken into account when assigning a weight. Therefore, we turn to heuristic methods. We use the \textit{Hamming distance}, standard measure of difference between binary arrays, defined for $\textbf{x, y}\in \mathbb{F}_2^n$ as 
 $h\left(\textbf{x, y}\right)=\sum_{i=1}^{n}x_i\neq y_i$, with the distance from the zero array called the \textit{Hamming weight}. Intuitively, as our target $\textbf{I}_{n}$ is a sparse matrix, we will prefer operations that minimize the Hamming weight of row operations. Even though the addition manifests itself as the binary XOR operation, we explore a wider choice of operations for the weight function and pick the best one as described below. Other operations might yield better results, since after tree selection a complex sequence of row additions is performed. We treat all the qubits symmetrically, therefore every entry in the vector will be calculated by the same rule, limiting the choice to all 2-bit operations. The usage of undirected graphs adds another constraint, 01 and 10 inputs must have the same output. A quick count shows that there are only $2^{3}=8$ candidates, including the irrelevant constant 0 function (ZERO) and the constant 1 function (ONE), which is equivalent to the unweighted case. More complex heuristic rules, such as bidirectional weights and parameterized scaling, did not affect the overall synthesis.

Eventually, given a parity matrix $M\in GL_{2}\left(n\right)$ and an edge $e=\left(u,v\right)$, the weight is given by
 \begin{equation}
     \label{weights}
    w_{f}\left(M, e=\left(u,v\right)\right)=h\left(f\left(M_{u},M_{v}\right)\right) 
\end{equation} 
where \textit{f} is one of the functions listed in \tref{tb: function table}. Therefore, when we come to use any linear synthesis algorithm, we have several options for the weighting heuristic. In order to compare the varying rules, we ``pre-train'' to find the ideal one, using a cost function estimating the average CNOT count of an algorithm utilizing a specific rule. Given a set of graphs $\mathcal{G}$ and a weight heuristic $w_{f}$ calculate 
\begin{equation} \label{eq:cost}
cost\left(f\right)=\sum_{G\in\mathcal{G}}\frac{1}{\left|G\right|^{2}}\left\langle CNOT\ count\left(G,M,w_{f}\right)\right\rangle _{M\in\mathcal{P}_{samp}\left(G\right)}
\end{equation}
where $\mathcal{P}_{samp}\left(G\right)$ is a set of random linear functions to synthesize on the graph G and $\left|G\right|$ is its cardinality. The function $CNOT\ count\left(G,M,w_{f}\right)$ returns the number of gates in the synthesized circuit and $\left\langle ...\right\rangle $ is used to average the count over all input matrices in $\mathcal{P}_{samp}\left(G\right)$. The idea is to find the best rule for different architectures, circuit widths, and linear functions. Normalization $\left|G\right|^{-2}$ is used to cancel the cost quadratic scaling in $\left|G\right|$, to provide a similar contribution from all graph sizes. 
\begin{table}[h!]
\caption{Two-bit operations used for the weight function and their corresponding cost estimation. 3 possible different inputs yielding a total of $2^3=8$ possible binary functions. The ZERO function assigns zero weight to all edges and is thus irrelevant for our purposes, and the ONE function is equivalent to an unweighted algorithm as all edges will hold the same weight. The cost function from Eq.~\eref{eq:cost} was evaluated by averaging synthesis over 100 random matrices for each of 6 graphs - grid and all-to-all architectures (low and high connectivity) with 9, 49, and 81 qubits. The uncertainty in the cost evaluation is 0.1 gates per qubit count squared for all entries.}
\label{tb: function table}
\begin{tabular}{@{}|l||l|l|l|l|l|l|l|l|}

\diagbox{input}{function}
 & ZERO & AND & XOR & OR & NOR & NXOR & NAND & ONE \\
\mr
00 & 0 & 0 & 0 & 0 & 1 & 1 & 1 & 1\\
01/10 & 0 & 0 & 1 & 1 & 0 & 0 & 1 & 1\\
11 & 0 & 1 & 0 & 1 & 0 & 1 & 0  & 1\\ 
\mr  
\RowCol cost & - & 4.1 & 3.6 & 3.6 & 3.8 & 3.8 & 3.5 & 3.7 \\ 
\SteinerGauss cost & - & 4.1 & 3.7 & 3.7 & 4.0 & 4.0 & 3.8 & 3.8 

\end{tabular}
\end{table}

 An illustration of the suggested heuristic is depicted in \tref{tb: Illustration}, in which we compare the steps of the augmented \RowCol algorithm with the NAND heuristic to the steps of the original \RowCol algorithm for a certain example. A similar weighting method has been proposed in \cite{Vandaele_2022} for phase polynomial synthesis, but differs from our work both in the problem it aims to solve and in the specific edge weighting. In their work, for each parity, they create a complete directed graph with the relevant qubits, then weight the edges according to $e=\left(u,v\right), M\in \mathbb{F}_2^{n\times m} \rightarrow w\left(e, M\right)=h\left(M_{u}\oplus M_{v}\right)-h\left(M_{v}\right)$, and find a minimal spanning tree.

\begin{table}
\caption{Illustration of the method. Operation of the RowCol algorithm on the same linear function and coupling map, with (left) and without (right) the heuristic. In each step of the process, the state of the parity matrix and the graph is presented, the terminal nodes are in gray and the chosen Steiner tree edges are colored in red. After eliminating the first 2 columns and rows, it is seen that the weighted algorithm has made larger progress, in terms of hamming distance from $\mathbb{I}_6$, and used smaller trees. The weighted algorithm synthesis ended with a circuit containing 18 CNOTs, 9 less than the unweighted one. For brevity, the remaining steps are shown in the appendix. 
}
\label{tb: Illustration}
\begin{tabular}{@{}l|ll|ll}
\toprule
Step & \multicolumn{2}{l}{Weighted} & \multicolumn{2}{|l}{Unweighted} \\ 
\mr
Col 0 & $\footnotesize \begin{pmatrix}
  1 & 0 & 0 & 1 & 1 & 0\\
  0 & 0 & 1 & 0 & 1 & 1\\
  0 & 1 & 1 & 1 & 1 & 1\\
  1 & 0 & 0 & 0 & 1 & 1\\
  1 & 1 & 0 & 0 & 1 & 1\\
  0 & 1 & 0 & 0 & 1 & 1\\
\end{pmatrix}$ & \includegraphics[width=8em]{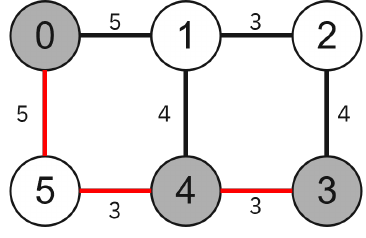} 
&
$\footnotesize \begin{pmatrix}
  1 & 0 & 0 & 1 & 1 & 0\\
  0 & 0 & 1 & 0 & 1 & 1\\
  0 & 1 & 1 & 1 & 1 & 1\\
  1 & 0 & 0 & 0 & 1 & 1\\
  1 & 1 & 0 & 0 & 1 & 1\\
  0 & 1 & 0 & 0 & 1 & 1\\
\end{pmatrix} $
&
\includegraphics[width=8em]{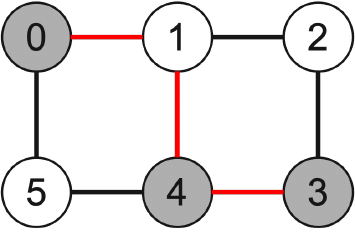}
\\
\mr
Row 0 & $\footnotesize \begin{pmatrix}
  1 & 0 & 0 & 1 & 1 & 0\\
  0 & 0 & 1 & 0 & 1 & 1\\
  0 & 1 & 1 & 1 & 1 & 1\\
  0 & 1 & 0 & 0 & 0 & 0\\
  0 & 1 & 0 & 0 & 1 & 1\\
  0 & 0 & 0 & 1 & 1 & 0\\
\end{pmatrix}$
& \includegraphics[width=8em]{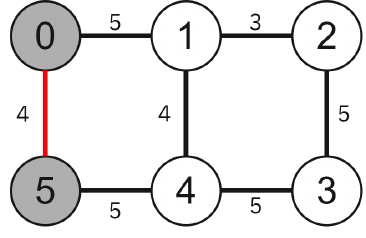} 
& 
$\footnotesize \begin{pmatrix}
  1 & 0 & 0 & 1 & 1 & 0\\
  0 & 1 & 1 & 1 & 1 & 0\\
  0 & 1 & 1 & 1 & 1 & 1\\
  0 & 1 & 0 & 0 & 0 & 0\\
  0 & 0 & 1 & 0 & 1 & 1\\
  0 & 1 & 0 & 0 & 1 & 1\\
\end{pmatrix}$ & 
\includegraphics[width=8em]{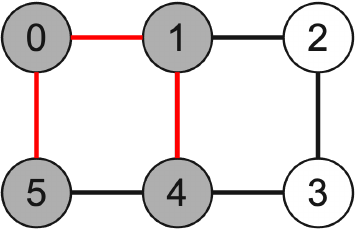} \\
\mr

Col 1 & $\footnotesize \begin{pmatrix}
  1 & 0 & 0 & 0 & 0 & 0\\
  0 & 0 & 1 & 0 & 1 & 1\\
  0 & 1 & 1 & 1 & 1 & 1\\
  0 & 1 & 0 & 0 & 0 & 0\\
  0 & 1 & 0 & 0 & 1 & 1\\
  0 & 0 & 0 & 1 & 1 & 0\\
\end{pmatrix}$
& \includegraphics[width=8em]{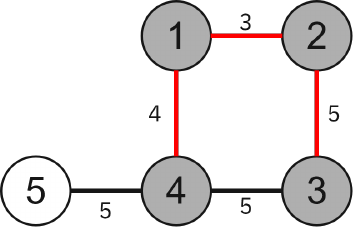} 
&
$\footnotesize \begin{pmatrix}
  1 & 0 & 0 & 0 & 0 & 0\\
  0 & 1 & 0 & 1 & 0 & 1\\
  0 & 1 & 1 & 1 & 1 & 1\\
  0 & 1 & 0 & 0 & 0 & 0\\
  0 & 0 & 1 & 0 & 1 & 1\\
  0 & 1 & 0 & 0 & 1 & 1\\
\end{pmatrix}$
&
\includegraphics[width=8em]{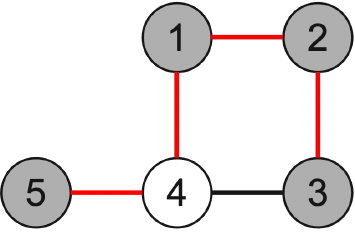} \\
\mr
Row 1 & $\footnotesize \begin{pmatrix}
  1 & 0 & 0 & 0 & 0 & 0\\
  0 & 1 & 1 & 0 & 0 & 0\\
  0 & 0 & 0 & 1 & 1 & 1\\
  0 & 0 & 1 & 1 & 1 & 1\\
  0 & 0 & 1 & 0 & 1 & 1\\
  0 & 0 & 0 & 1 & 1 & 0\\
\end{pmatrix}$
& \includegraphics[width=8em]{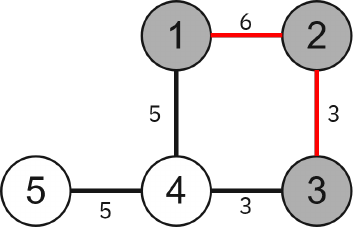} 
&
$\footnotesize \begin{pmatrix}
  1 & 0 & 0 & 0 & 0 & 0\\
  0 & 1 & 1 & 0 & 0 & 0\\
  0 & 0 & 0 & 1 & 1 & 1\\
  0 & 0 & 1 & 1 & 1 & 1\\
  0 & 0 & 1 & 0 & 1 & 1\\
  0 & 0 & 0 & 1 & 1 & 0\\
\end{pmatrix}$
&
\includegraphics[width=8em]{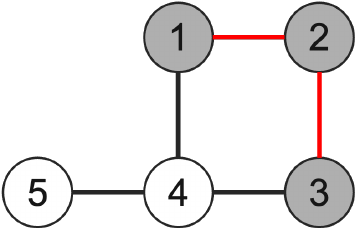} \\
\mr
Col 2 & $\footnotesize \begin{pmatrix}
  1 & 0 & 0 & 0 & 0 & 0\\
  0 & 1 & 0 & 0 & 0 & 0\\
  0 & 0 & 1 & 0 & 0 & 0\\
  0 & 0 & 1 & 1 & 1 & 1\\
  0 & 0 & 1 & 0 & 1 & 1\\
  0 & 0 & 0 & 1 & 1 & 0\\
\end{pmatrix}$
& \includegraphics[width=8em]{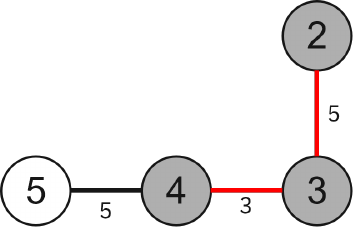} 
&
$\footnotesize \begin{pmatrix}
  1 & 0 & 0 & 0 & 0 & 0\\
  0 & 1 & 0 & 0 & 0 & 0\\
  0 & 0 & 0 & 1 & 0 & 1\\
  0 & 0 & 1 & 1 & 1 & 1\\
  0 & 0 & 1 & 1 & 0 & 1\\
  0 & 0 & 1 & 0 & 1 & 1\\
\end{pmatrix}$
&
\includegraphics[width=8em]{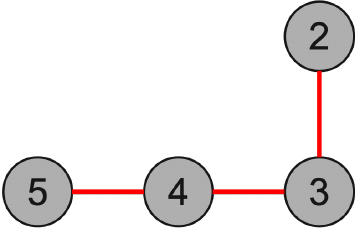}
\end{tabular}
\end{table} 

In order to update the weights, after each iteration, one should traverse all edges of the coupling map and perform an operation with a time complexity of $\mathcal{O}\left(\left|V\right|\right)$. Consequently, the resulting complexity overhead of the heuristic is a polynomial addition of $\left(\left|V\right|\cdot\left|E\right|\right)$ for a coupling graph with $\left|V\right|$ nodes and $\left|E\right|$ edges. To continue with the \RowCol example, for an n-qubit circuit, the algorithm performs n iterations, the computationally heaviest operation in each being the Steiner tree search. Given that the approximate tree search complexity is $\mathcal{O}\left(\left|S\right|\left(\left|V\right| \log\left|V\right| + \left|E\right|\right)\right)$\cite{MEHLHORN1988125}, the enhanced program asymptotic runtime is 
\begin{equation}
\mathcal{O}\left[\underset{\#iterations}{\underbrace{\left|V\right|}}\cdot\left(\underset{wt. track}{\underbrace{\left|V\right|\cdot\left|E\right|}}+\underset{SteinerTree}{\underbrace{\left|S\right|\left(\left|V\right|\log\left|V\right|+\left|E\right|\right)}}\right)\right]=\mathcal{O}\left(\left|V\right|^4\right)
\end{equation}
which is equal to the complexity of the original \RowCol algorithm. 

\section{Results}
\label{sec:experiments}
\begin{figure}[h!]
\centering
    \begin{subfigure}[t]{0.48\textwidth}
         \centering
         \includegraphics[width=\textwidth]{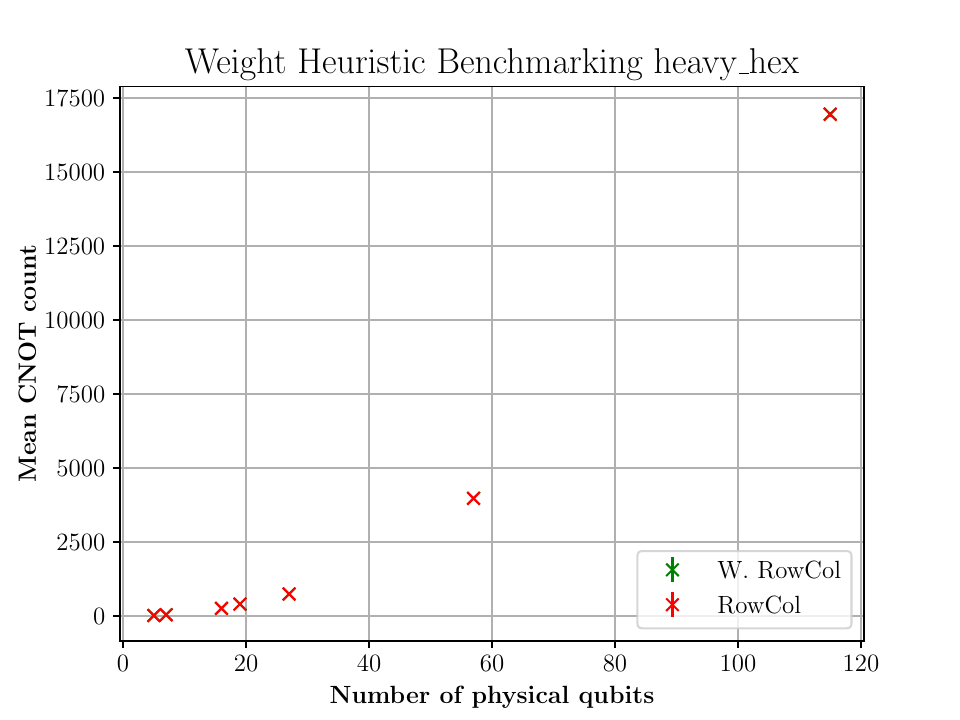}
         \caption{heavy-hex architecture. A decreasing Steiner tree is not always feasible in these graphs, therefore \SteinerGauss is not evaluated. In this case, the weighted and unweighted points are indistinguishable.}
         \label{fig:hhex}
      \end{subfigure}
     \begin{subfigure}[t]{0.48\textwidth}
         \centering
         \includegraphics[width=\textwidth]{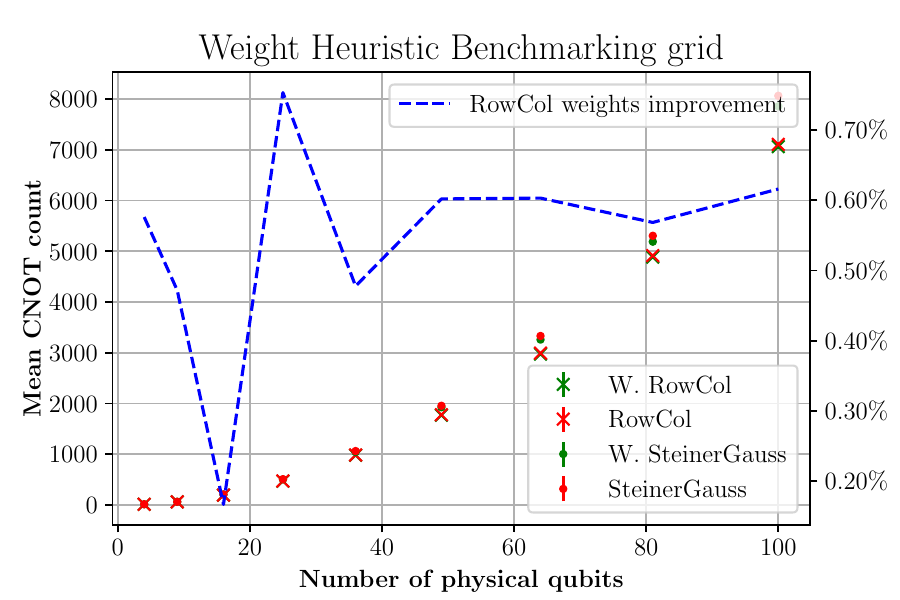}
         \caption{2D grid}
         \label{fig:grid}
     \end{subfigure}
     \hfill
     \begin{subfigure}[t]{0.48\textwidth}
         \centering
         \includegraphics[width=\textwidth]{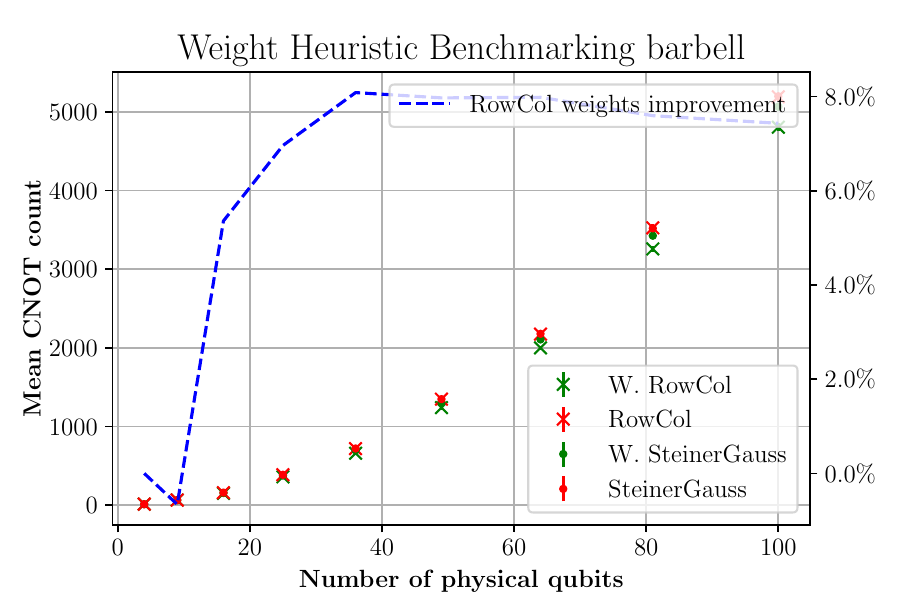}
         \caption{barbell, a graph containing 2 complete subgraphs connected by a path of 1 or 2 edges}
         \label{fig:barbell}
     \end{subfigure}
     \hfill
     \begin{subfigure}[t]{0.48\textwidth}
         \centering
         \includegraphics[width=\textwidth]{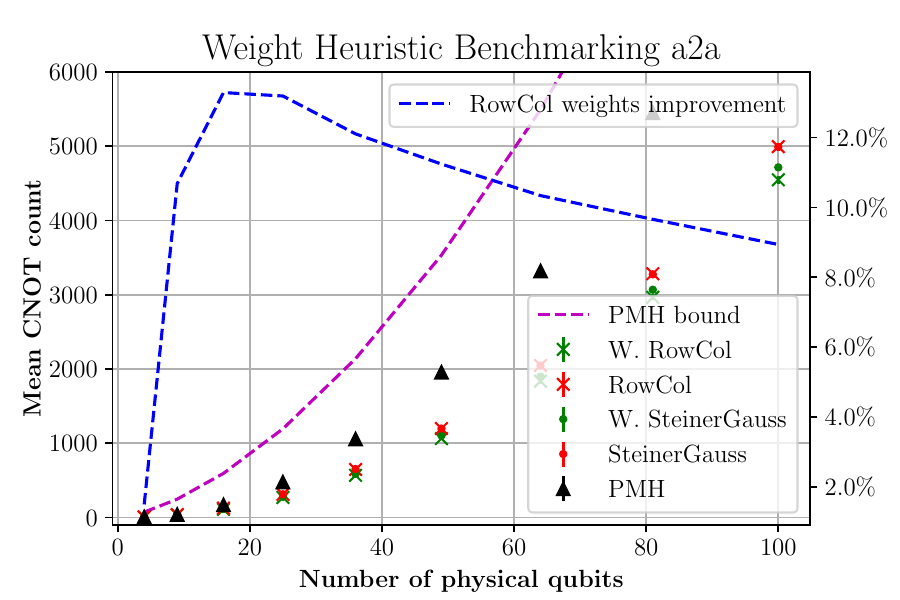}
         \caption{All-to-All connectivity, with Patel-Markov-Hayes(PMH)\cite{PMH} results  as comparison. PMH result for 100 qubits is 8175 $\pm$ 27, and has been omitted for clarity}
         \label{fig:a2a}
     \end{subfigure}
    \caption{Benchmarking. Each marker represents an average result over one hundred reversible matrices of the same size, standard deviations are around tens of gates thus not visible. Markers are grouped by algorithm (symbol) and weights (color). The weighted \protect\RowCol algorithm uses the NAND heuristic, while the \protect\SteinerGauss results shown are with the OR rule. The blue dashed line together with the right axis represents the weighted version improvement, calculated by $1-\left\langle \frac{CNOT\text{count}\left(G,M,w_\text{NAND}\right)}{CNOT\text{count}\left(G,M,w_\text{const}\right)}\right\rangle$.}
    \label{fig:benchmarking}
\end{figure}

 We now turn to present the performance results of the method for the two different algorithms \RowCol and \SteinerGauss. As described in the Method section, we first looked for the optimal weight rule by minimizing the cost function in \eref{eq:cost}. In this "pre-training" process, we found the NAND rule optimal for \RowCol with an average cost of 3.5 gates per qubit number squared, and for the \SteinerGauss algorithm the OR heuristic yielded the optimal result, as can be seen in the bottom part of \tref{tb: function table}. Intuitively, since in the \RowCol algorithm, one performs row eliminations by linear combinations of other rows, reducing the number of dependent rows will lead to better results. Therefore, the NAND operation, which favors cancellation of the overlapping entries 1, was found to perform best. In \fref{fig:benchmarking} we depict the average CNOT count of the algorithms, for various device connectivity architectures and sizes, showing a substantial improvement for the weighted algorithms.  
 
\begin{figure}[h!]
\centering
     \begin{subfigure}[t]{0.48\textwidth}
         \centering
         \includegraphics[width=\textwidth]{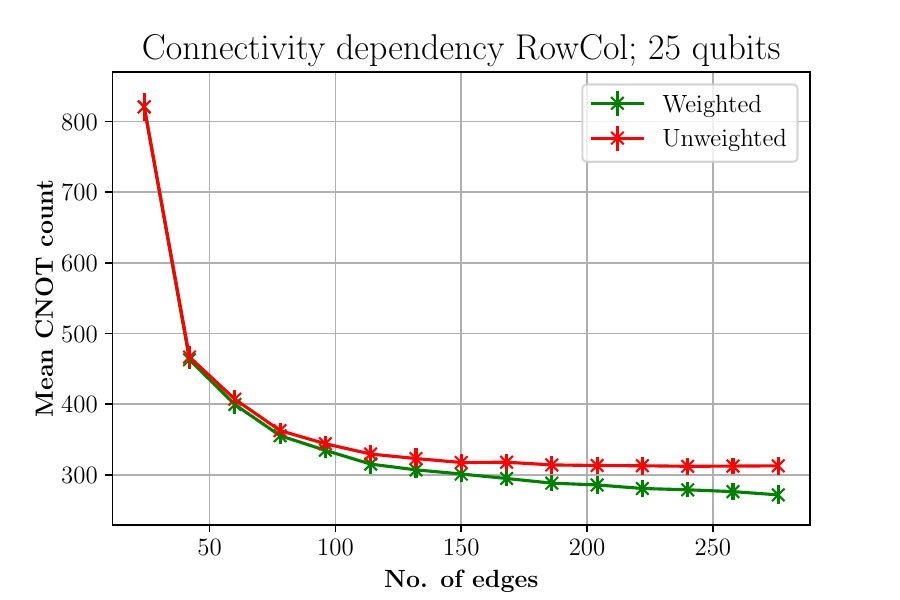}
         \caption{size}
         \label{fig:edge_size}
     \end{subfigure}
     \hfill
     \begin{subfigure}[t]{0.48\textwidth}
         \centering
         \includegraphics[width=\textwidth]{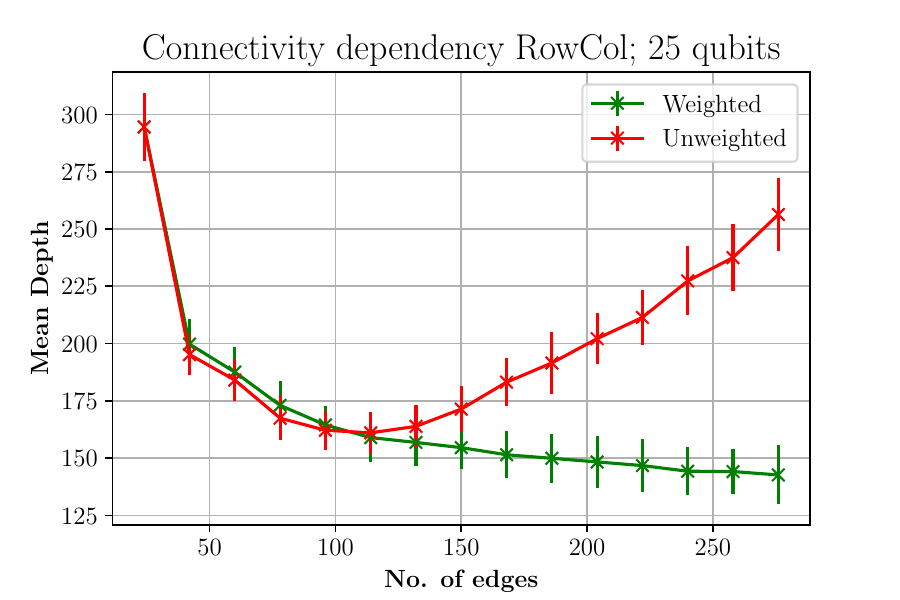}
         \caption{depth}
         \label{fig:edge_depth}
     \end{subfigure}
        \caption{synthesis performance as a function of the coupling map connectivity}
        \label{fig:edges}
    \hfill
\end{figure}

\begin{figure}[h!]
\centering
     \begin{subfigure}[t]{0.48\textwidth}
         \centering
         \includegraphics[width=\textwidth]{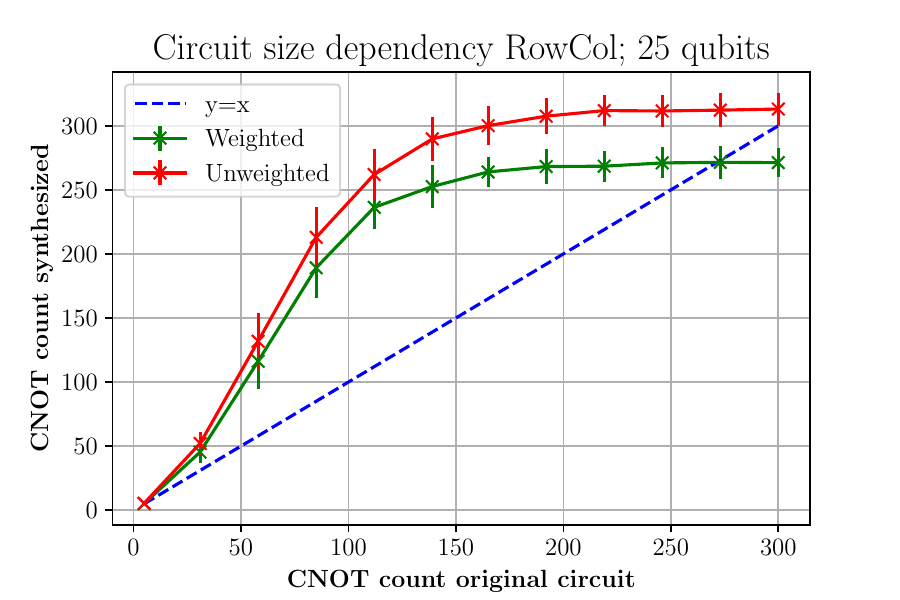}
         \caption{size}
         \label{fig:inputsize_size}
     \end{subfigure}
     \hfill
     \begin{subfigure}[t]{0.48\textwidth}
         \centering
         \includegraphics[width=\textwidth]{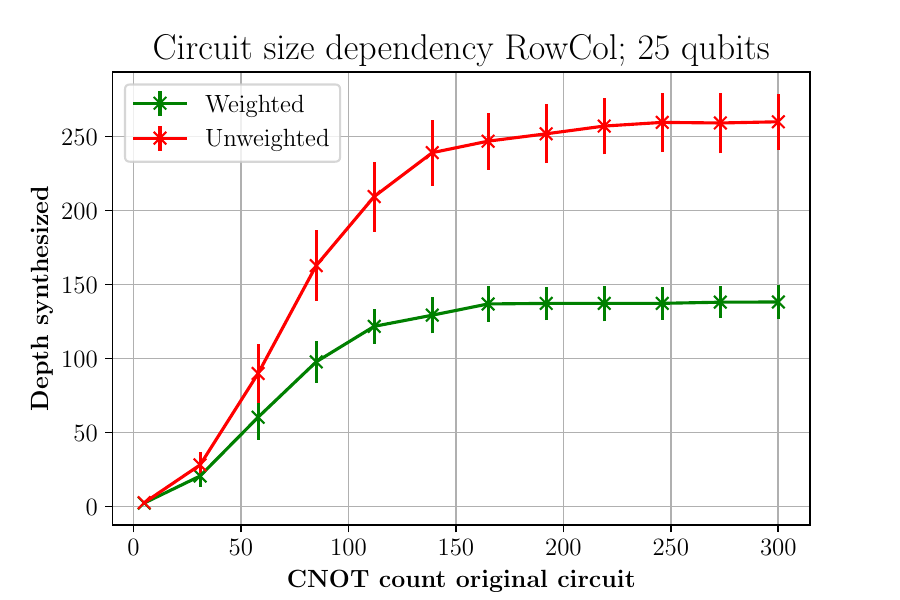}
         \caption{depth}
         \label{fig:inputsize_depth}
     \end{subfigure}
        \caption{input circuit CNOT count dependency. In all of these simulations, the coupling map is fixed to be a complete graph with 25 qubits. Each point represents the average result over 100 random parity matrices with the same amount of row operations (sometimes annulling) on the identity. }
        \label{fig:input_size}
    \hfill
\end{figure}

Since the method requires enough Steiner tree possibilities within the graph to make a difference, the more connected the device, the larger the circuit size reduction. Our results show a negligible change for the heavy-hex architecture which is within the error bounds, a slight 1\% boost given a grid of qubits, and up to more than 10\% smaller circuits on complete graphs. To quantify this claim, in \fref{fig:edge_size} we present the synthesis performance as a function of the connectivity of the graph. On a (fixed) 25-qubit device initially containing only chain nearest neighbor edges, we compare the synthesis of the same matrices as we add randomly selected connecting edges. Clearly, the weighted algorithm uses the higher connectivity more intelligently. The improvement is even greater when we look at the depth metric within the simulation mentioned, as seen in \fref{fig:edge_depth}. We explain this gap by the fact that during the unweighted algorithm the lowest (by index) Steiner nodes are chosen repeatedly, while the weight analysis forces a more diverse choice, which opens the possibility for parallel gates and lower overall synthesis depth. 
Naturally, the performance of the algorithms depends on the input linear circuit properties as well. We have trialed random input circuits with an increasing number of CNOTs and analyzed the properties of the re-synthesized circuits. From \fref{fig:inputsize_size} (\ref{fig:inputsize_depth}) we infer that the more CNOTs in the input circuits, the larger the reduction in the size (depth) of the output circuits. This gain reaches saturation since the $GL_2 (n)$ group contains a finite number of unique operations, many of the two-qubit operations in input circuits of large size cancel each other out, and the average synthesis cost approaches a constant value. 

\section{Conclusion}
\label{sec:conclusion}
In this paper, we present a general heuristic to improve Steiner tree algorithms. The proposed approach has been tested on the \RowCol and \SteinerGauss algorithms, improving their performance on varying architectures and linear functions. Our benchmarking showed that the heuristic is most useful in cases where the graphs are highly connected or where large circuits need to be synthesized. In addition, the complexity overhead of the heuristic is minimal, making it suitable for real-world applications. This work lays the foundation for future algorithms that can be built upon the proposed heuristic, adjusting the assignment of the weights to new optimization tasks and methods in the field. 

\nonumsection{Acknowledgements}
The authors thank Lev Bishop, Eli Arbel, Simon Martiel, Ali Javadi-Abhari, and Dmitri Maslov from IBM Quantum for fruitful discussions and support. We also thank Arianne Meijer - van de Griend for the code and explanation of the \SteinerGauss algorithm. 

    \bibliographystyle{unsrt}
    \bibliography{Weighted_Steiner}

\newpage
\appendix { End of Illustration}
\label{illustrationAppendix}

In this appendix, we present the remaining elimination steps of the example presented in the main text, depicted in \tref{tb: IllustrationAppendixTable}. The resulting circuits of this synthesis example appear in \fref{fig:circs}
\begin{table}[h!]
\caption{Remaining steps of the example from \tref{tb: Illustration}}
\label{tb: IllustrationAppendixTable}
\begin{tabular}{@{}l|ll|ll}
\toprule
Step & \multicolumn{2}{l}{Weighted} & \multicolumn{2}{|l}{Unweighted} \\ 
\midrule
Row 2 & $ \footnotesize \begin{pmatrix}
  1 & 0 & 0 & 0 & 0 & 0\\
  0 & 1 & 0 & 0 & 0 & 0\\
  0 & 0 & 1 & 0 & 0 & 0\\
  0 & 0 & 0 & 1 & 1 & 1\\
  0 & 0 & 0 & 1 & 0 & 0\\
  0 & 0 & 0 & 1 & 1 & 0\\
\end{pmatrix}$ & \includegraphics[width=9em]{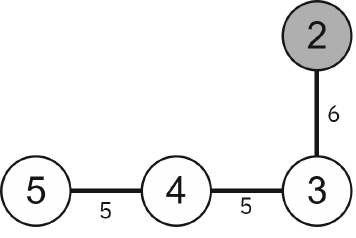} 
&
$ \footnotesize \begin{pmatrix}
  1 & 0 & 0 & 0 & 0 & 0\\
  0 & 1 & 0 & 0 & 0 & 0\\
  0 & 0 & 1 & 0 & 1 & 0\\
  0 & 0 & 0 & 1 & 0 & 1\\
  0 & 0 & 0 & 0 & 1 & 0\\
  0 & 0 & 0 & 1 & 1 & 0\\
\end{pmatrix}$
&
\includegraphics[width=9em]{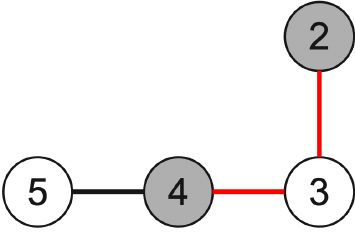}
\\ 
\mr
Col 3 & $ \footnotesize \begin{pmatrix}
  1 & 0 & 0 & 0 & 0 & 0\\
  0 & 1 & 0 & 0 & 0 & 0\\
  0 & 0 & 1 & 0 & 0 & 0\\
  0 & 0 & 0 & 1 & 1 & 1\\
  0 & 0 & 0 & 1 & 0 & 0\\
  0 & 0 & 0 & 1 & 1 & 0\\
\end{pmatrix}$ 
& \includegraphics[width=9em]{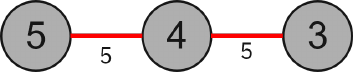} 
& 
$ \footnotesize \begin{pmatrix}
  1 & 0 & 0 & 0 & 0 & 0\\
  0 & 1 & 0 & 0 & 0 & 0\\
  0 & 0 & 1 & 0 & 0 & 0\\
  0 & 0 & 0 & 1 & 1 & 1\\
  0 & 0 & 0 & 0 & 1 & 0\\
  0 & 0 & 0 & 1 & 1 & 0\\
\end{pmatrix}$ & 
\includegraphics[width=9em]{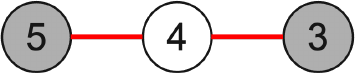} \\
\mr
Row 3 & $ \footnotesize \begin{pmatrix}
  1 & 0 & 0 & 0 & 0 & 0\\
  0 & 1 & 0 & 0 & 0 & 0\\
  0 & 0 & 1 & 0 & 0 & 0\\
  0 & 0 & 0 & 1 & 1 & 1\\
  0 & 0 & 0 & 0 & 1 & 1\\
  0 & 0 & 0 & 0 & 1 & 0\\
\end{pmatrix}$
& \includegraphics[width=9em]{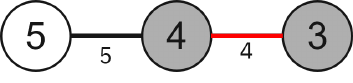} 
&
$\footnotesize \begin{pmatrix}
  1 & 0 & 0 & 0 & 0 & 0\\
  0 & 1 & 0 & 0 & 0 & 0\\
  0 & 0 & 1 & 0 & 0 & 0\\
  0 & 0 & 0 & 1 & 1 & 1\\
  0 & 0 & 0 & 0 & 1 & 1\\
  0 & 0 & 0 & 0 & 1 & 0\\
\end{pmatrix}$
&
\includegraphics[width=9em]{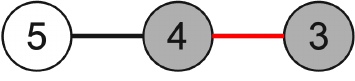} \\
\mr
Col 4 & $ \footnotesize \begin{pmatrix}
  1 & 0 & 0 & 0 & 0 & 0\\
  0 & 1 & 0 & 0 & 0 & 0\\
  0 & 0 & 1 & 0 & 0 & 0\\
  0 & 0 & 0 & 1 & 0 & 0\\
  0 & 0 & 0 & 0 & 1 & 1\\
  0 & 0 & 0 & 0 & 1 & 0\\
\end{pmatrix}$
& \includegraphics[width=9em]{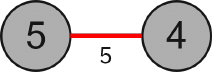} 
&
$ \footnotesize \begin{pmatrix}
  1 & 0 & 0 & 0 & 0 & 0\\
  0 & 1 & 0 & 0 & 0 & 0\\
  0 & 0 & 1 & 0 & 0 & 0\\
  0 & 0 & 0 & 1 & 0 & 0\\
  0 & 0 & 0 & 0 & 1 & 1\\
  0 & 0 & 0 & 0 & 1 & 0\\
\end{pmatrix}$
&
\includegraphics[width=9em]{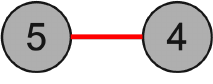} \\
\mr
Row 4 & $\footnotesize \begin{pmatrix}
  1 & 0 & 0 & 0 & 0 & 0\\
  0 & 1 & 0 & 0 & 0 & 0\\
  0 & 0 & 1 & 0 & 0 & 0\\
  0 & 0 & 0 & 1 & 0 & 0\\
  0 & 0 & 0 & 0 & 1 & 1\\
  0 & 0 & 0 & 0 & 0 & 1\\
\end{pmatrix}$
& \includegraphics[width=9em]{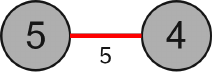} 
&
$\footnotesize  \begin{pmatrix}
  1 & 0 & 0 & 0 & 0 & 0\\
  0 & 1 & 0 & 0 & 0 & 0\\
  0 & 0 & 1 & 0 & 0 & 0\\
  0 & 0 & 0 & 1 & 0 & 0\\
  0 & 0 & 0 & 0 & 1 & 1\\
  0 & 0 & 0 & 0 & 0 & 1\\
\end{pmatrix}$
&
\includegraphics[width=9em]{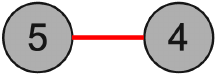}
\end{tabular}
\end{table}

\begin{figure}
    \centering
    \begin{subfigure}[b]{\textwidth}
         \centering
         \Qcircuit @C=1.0em @R=0.8em @!R { \\
	 	\nghost{{q}_{0} :  } & \lstick{{q}_{0} :  } & \qw & \qw & \qw & \qw & \qw & \qw & \targ & \qw & \ctrl{5} & \qw & \qw & \qw & \qw & \qw & \qw & \qw & \qw & \qw\\
	 	\nghost{{q}_{1} :  } & \lstick{{q}_{1} :  } & \qw & \qw & \qw & \qw & \qw & \targ & \qw & \qw & \qw & \ctrl{1} & \qw & \ctrl{3} & \targ & \qw & \qw & \qw & \qw & \qw\\
	 	\nghost{{q}_{2} :  } & \lstick{{q}_{2} :  } & \qw & \qw & \qw & \qw & \ctrl{1} & \ctrl{-1} & \qw & \targ & \qw & \targ & \ctrl{1} & \qw & \qw & \qw & \qw & \qw & \qw & \qw\\
	 	\nghost{{q}_{3} :  } & \lstick{{q}_{3} :  } & \qw & \qw & \targ & \ctrl{1} & \targ & \ctrl{1} & \qw & \ctrl{-1} & \qw & \qw & \targ & \qw & \qw & \qw & \targ & \qw & \qw & \qw\\
	 	\nghost{{q}_{4} :  } & \lstick{{q}_{4} :  } & \targ & \ctrl{1} & \ctrl{-1} & \targ & \ctrl{1} & \targ & \qw & \qw & \qw & \qw & \qw & \targ & \ctrl{-3} & \targ & \ctrl{-1} & \ctrl{1} & \qw & \qw\\
	 	\nghost{{q}_{5} :  } & \lstick{{q}_{5} :  } & \ctrl{-1} & \targ & \qw & \qw & \targ & \qw & \ctrl{-5} & \qw & \targ & \qw & \qw & \qw & \qw & \ctrl{-1} & \qw & \targ & \qw & \qw\\
\\ } 
         \caption{weighted}
         \label{fig:we_circ}
     \end{subfigure}
     \hfill
     \begin{subfigure}[b]{\textwidth}
         \centering
         \makebox[\textwidth]{
        \Qcircuit @C=1.0em @R=0.8em @!R { \\
	 	\nghost{{q}_{0} :  } & \lstick{{q}_{0} :  } & \qw & \qw & \qw & \qw & \qw & \qw & \qw & \qw & \qw & \qw & \qw & \qw & \qw & \qw & \qw & \qw & \targ & \qw & \qw & \targ & \ctrl{1} & \qw & \qw & \qw & \qw & \qw\\
	 	\nghost{{q}_{1} :  } & \lstick{{q}_{1} :  } & \qw & \qw & \qw & \qw & \qw & \qw & \qw & \qw & \qw & \qw & \qw & \targ & \qw & \ctrl{1} & \qw & \ctrl{3} & \ctrl{-1} & \qw & \targ & \qw & \targ & \ctrl{3} & \qw & \targ & \qw & \qw\\
	 	\nghost{{q}_{2} :  } & \lstick{{q}_{2} :  } & \qw & \qw & \qw & \qw & \targ & \qw & \qw & \targ & \ctrl{1} & \qw & \targ & \ctrl{-1} & \targ & \targ & \ctrl{1} & \qw & \qw & \qw & \qw & \qw & \qw & \qw & \qw & \qw & \qw & \qw\\
	 	\nghost{{q}_{3} :  } & \lstick{{q}_{3} :  } & \qw & \qw & \targ & \ctrl{1} & \ctrl{-1} & \qw & \targ & \ctrl{-1} & \targ & \ctrl{1} & \ctrl{-1} & \qw & \ctrl{-1} & \qw & \targ & \qw & \qw & \qw & \qw & \qw & \qw & \qw & \targ & \qw & \qw & \qw\\
	 	\nghost{{q}_{4} :  } & \lstick{{q}_{4} :  } & \targ & \ctrl{1} & \ctrl{-1} & \targ & \ctrl{1} & \targ & \ctrl{-1} & \qw & \qw & \targ & \ctrl{1} & \qw & \qw & \qw & \qw & \targ & \ctrl{1} & \targ & \ctrl{-3} & \qw & \qw & \targ & \ctrl{-1} & \ctrl{-3} & \qw & \qw\\
	 	\nghost{{q}_{5} :  } & \lstick{{q}_{5} :  } & \ctrl{-1} & \targ & \qw & \qw & \targ & \ctrl{-1} & \qw & \qw & \qw & \qw & \targ & \qw & \qw & \qw & \qw & \qw & \targ & \ctrl{-1} & \qw & \ctrl{-5} & \qw & \qw & \qw & \qw & \qw & \qw }}
         \caption{unweighted}
         \label{fig:uw_circ}
     \end{subfigure}
    \caption{Resulting circuits}
    \label{fig:circs}
\end{figure}

\end{document}